\documentclass[a4paper,11pt]{article}

\usepackage[utf8]{inputenc}
\usepackage[T1]{fontenc}
\usepackage{amsmath}
\usepackage{amsfonts}
\usepackage{amssymb}
\usepackage[english]{babel}
\usepackage{float}
\usepackage{graphicx}
\usepackage[rmargin=2.5cm,lmargin=2.5cm,tmargin=3cm]{geometry}
\usepackage{lmodern}
\usepackage{url}
\usepackage{hyperref}
\usepackage{lastpage}
\usepackage{lettrine}
\usepackage{color}
\usepackage{tikz}
\usepackage{tikzscale}
\usetikzlibrary{calc, arrows, backgrounds, trees}
\usepackage{pdflscape}
\usepackage[nottoc]{tocbibind}
\usepackage{booktabs}
\usepackage{arydshln}
\usepackage[export]{adjustbox}
\usepackage{float}
\usepackage{rotating}
\usepackage{nameref}
\usepackage{stmaryrd}
\usepackage{verbatim}
\usepackage{color, colortbl}
\usepackage{makecell}

\usepackage{amsmath}
\usepackage{amsthm}
\usepackage{mathtools}

\usepackage{natbib}
\setcitestyle{aysep={}} 
\usepackage{url}

\newtheorem{theorem}{Theorem}
\newtheorem{lemma}{Lemma}

\newtheorem{corollary}{Corollary}

\graphicspath{{Figures/}}       			

\usepackage{subcaption}

\usepackage{authblk}
\newcommand{\keywords}[1]{{\small{\noindent \bfseries Keywords: }#1}} 

\title{A note on the complexity of the picker routing problem in multi-block warehouses and related problems}
\date{}

\author[1,2]{Thibault Prunet}
\author[1,3]{Nabil Absi}
\author[4,5]{Diego Cattaruzza}
\affil[1]{\small Mines Saint-Etienne, Univ. Clermont Auvergne, INP Clermont Auvergne, CNRS UMR 6158 LIMOS, F-13120 Gardanne, France, {absi@emse.fr}}
\affil[2]{\small CERMICS, Ecole des Ponts, IP Paris, Marne-la-Vallée, France, {thibault.prunet@enpc.fr}}
\affil[3]{\small SOLA Center, Africa Business School, Mohammed VI Polytechnic University, Rabat,
Morocco}
\affil[4]{\small CRIStAL Centre de Recherche en Informatique Signal et Automatique de Lille, University Lille, CNRS, Centrale Lille, Inria, UMR 9189, F-59000 Lille, France}
\affil[5]{\small Department of Mathematics, Computer Science and Physics, University of Udine, Via delle Scienze 206, 33100 Udine, Italy {diego.cattaruzza@uniud.it}}

\begin{document}

\maketitle

\begin{abstract}
\noindent
The Picker Routing Problem (PRP), which consists of finding a minimum-length tour between a set of storage locations in a warehouse, is one of the most important problems in the warehousing logistics literature. Despite its popularity, the tractability of the PRP in multi-block warehouses remains an open question. This technical note aims to fill this research gap by establishing that the problem is strongly NP-hard. As a corollary, the complexity status of other related problems is settled.

\end{abstract}



\section{Introduction}

The Picker Routing Problem (PRP), also called the single picker routing problem or the order picking problem, is one of the most important and most studied problems in the warehousing logistics literature \citep{de_koster_design_2007,gu_research_2007}. Indeed, the process of retrieving products from their storage locations to serve customer orders, known as Order Picking (OP), is classically considered the most resource-intensive operation in warehouse management. According to \cite{tompkins_facilities_2010}, OP accounts for 55\% of the total operating cost in manual picker-to-parts warehouses, with travel time accounting for more than 50\% of the picking time. It is therefore not surprising that the PRP generated significant attention in the literature, both as a stand-alone problem \citep{masae_order_2020}, and when integrated with other planning problems \citep{van_gils_designing_2018}. In their literature review, \cite{masae_order_2020} identified 203 journal papers on the PRP since the seminal work of \cite{ratliff_order-picking_1983}. The research stream remains highly active, as more than 50\% of this corpus was published in the five-year period prior to their work.

In this note, we focus on the PRP on a rectangular warehouse with parallel picking aisles, grouped by \textit{blocks} separated by a number of cross aisles. Figure~\ref{fig:layout_2blocks} provides an illustration of such a warehouse with two blocks. The recent review of \cite{masae_order_2020} also highlights that parallel-aisle warehouses are used in the vast majority of the related literature. 

Despite its popularity and importance in the warehousing literature, the tractability of the PRP in conventional warehouses remains an open question. A first attempt to answer this question was carried out by \cite{celik_order_2019}, who proved the intractability of the problem when distances are not constant between consecutive aisles and locations. It is important to emphasize that, despite the theoretical interest of this result, it does not provide insights about the complexity of the PRP on conventional warehouses that represent the vast majority of the literature on the topic. This gap is the focus of the present work, which is structured as follows. We define the problem in Section~\ref{sec:definition}, review related complexity results from the literature in Section~\ref{sec:literature} and prove the strong NP-hardness of the PRP in Section~\ref{sec:complexity}, use this result to prove the intractability of other related problems in Section~\ref{sec:additional} and conclude with Section~\ref{sec:conclusion}.


\section{Problem definition}\label{sec:definition}

\begin{figure}[h!]
    \centering
    \includegraphics{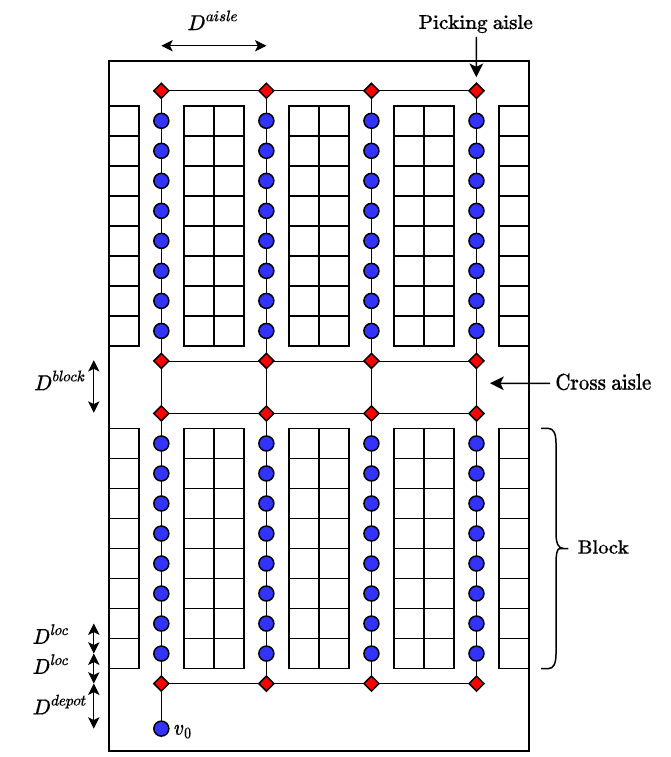}
    \caption{Example of a two-block layout. Blue dots represent locations, and red diamonds represent dummy points corresponding to intersections between aisles and cross aisles.}
    \label{fig:layout_2blocks}
\end{figure}

The PRP is defined as follows. We consider a warehouse layout composed of a set $\mathcal{K} = \{1,\dots,K\}$ of blocks, and a set $\mathcal{A} = \{1,\dots,A\}$ of aisles in each block. We assume pickers can pick indifferently from the right and left shelves when crossing an aisle. Therefore, we can group equivalent storage spaces in terms of distances (i.e., the ones facing each other), referring to them as {\em location}. 
In this case, we note $\mathcal{L}^{ka} = \{l^{ka}_i,\, 1 \leq i \leq L\}$ the set of locations in block $k \in \mathcal{K}$, aisle $a \in \mathcal{A}$, and $\mathcal{L} = \bigcup_{k \in \mathcal{K}, a \in \mathcal{A}} \mathcal{L}^{ka}$ represents the set of all locations. We note $D^{loc}$ the distance between two consecutive locations, $D^{aisle}$ the distance between two consecutive aisles, and $D^{block}$ the distance between two consecutive blocks. Additionally, we assume that a picking tour starts and ends at the same location $v_0$ called the {\em depot}. The depot is located in the first cross aisle, in front of aisle $a^{depot} \in \mathcal{A}$, and separated by a distance $D^{depot}$ from the beginning of aisle $a^{depot}$. Figure~\ref{fig:layout_2blocks} provides an illustration with the associated notation. 

A subset $\mathcal{V} \subset \mathcal{L}$ of locations needs to be visited, and we note $\mathcal{V}^{PRP} = \mathcal{V} \cup \{v_0\}$. The PRP is then defined on the complete undirected graph $\mathcal{G}^{PRP} = (\mathcal{V}^{PRP},\mathcal{E}^{PRP})$, where each vertex corresponds to a location to be visited. Each edge $e = (v^1,v^2) \in \mathcal{E}^{PRP}$ is associated with a weight $c_e$, corresponding to the shortest walking distance between locations $v^1$ and $v^2$. The objective of the PRP is to find a tour of minimum weight in $\mathcal{G}^{PRP}$ that visits all vertices exactly once.


\section{Previous complexity results}\label{sec:literature}

\begin{figure}[h!]
    \centering
    \includegraphics[width=0.8\textwidth]{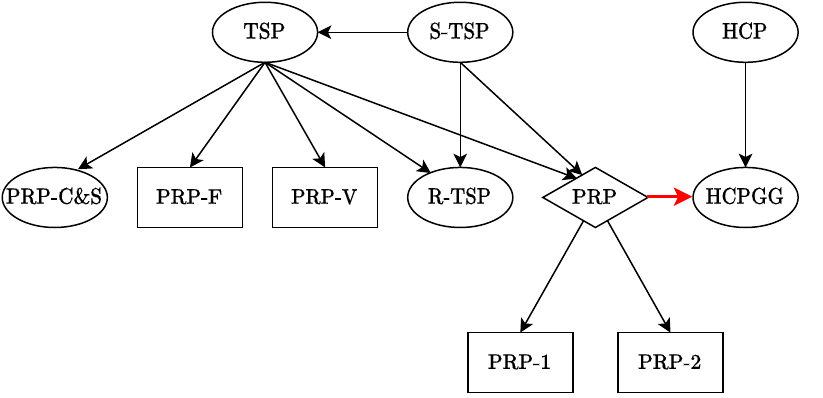}
    \caption{Complexity map of the problems related to the PRP. Round nodes represent NP-hard problems (or NP-complete problems for decision problems), while square nodes represent polynomially solvable problems. Diamonds represent problems whose tractability is undetermined. An arrow represents a link of polynomial reducibility between two problems, where the problem at its destination is reducible to the problem at its origin. For instance, the PRP is reducible to the TSP, as TSP is a generalization of the PRP. Note that the polynomial reducibility is a transitive relationship. The red arrow is the reduction we use in the proof of Theorem~\ref{theorem:complexity}.}
    \label{fig:complexity_map}
\end{figure}

In this section, we review the results of complexity known in the literature concerning problems related to the PRP. Figure~\ref{fig:complexity_map} provides a graphical representation of the relationship between the mentioned problems. It is important to note that Figure~\ref{fig:complexity_map} is not exhaustive, but serves as a comprehensive illustration for this section. An interested reader is referred to \cite{celik_order_2019} for a more complete review of the topic.

\paragraph{Known complexity results for PRP variants.}

The PRP defined on a single block layout (PRP-1) is known to be polynomially solvable since the seminal work of \cite{ratliff_order-picking_1983}, which introduced a dynamic programming-based algorithm to solve the problem optimally. Furthermore, it is notable that the complexity of the algorithm of \cite{ratliff_order-picking_1983} is linear in the number of locations to be visited and the number of aisles in the layout, that is $\mathcal{O}(|\mathcal{V}^{PRP}| + A)$ \citep{hesler_note_2022}. This result was later extended by \cite{roodbergen_routing_2001-1} to the case of a two-block warehouse layout, referred to as PRP-2. The recent work of \cite{cambazard_fixed-parameter_2018} on the Rectilinear Traveling Salesman Problem (R-TSP) brought a new perspective to the topic. They propose a fixed-parameter algorithm for the R-TSP, subsequently adapting it to the multi-block PRP in their follow-up study \citep{pansart_exact_2018}. Interestingly, their algorithm has a complexity of $\mathcal{O}(|\mathcal{V}^{PRP}| + KA7^K)$, which is only superlinear in the number of blocks $K$. However, to the best of the authors' knowledge, the complexity of the PRP with an unbounded number of blocks remains unknown, which is surprising considering the importance of the problem in the field of warehousing logistics.

The work of \cite{celik_order_2019} is the sole attempt known by the authors to study the complexity of the PRP. In their study, they prove the NP-hardness of the PRP variant, noted PRP-C\&S, where distances between consecutive locations and consecutive aisles are not constant. According to the authors, \lq\lq most, if not all instances of the [PRP] involve equal travel times between two consecutive pick aisles and equal subaisle lengths. In such a case, the transformation in the proof [...] ceases to be polynomial\rq\rq. It is important to emphasize that the PRP-C\&S is a less structured relaxation of the PRP studied in the current paper and does not correspond to conventional layouts that reflect the majority of the warehouses considered in the literature. If we consider other unconventional warehouse layouts, which represent a small minority of the works on the topic, \cite{celk_order_2014} propose a polynomial algorithm for the PRP on the so-called \textit{fishbone} and \textit{flying-V} layouts, where the lengths and orientations of the picking aisles are no longer constant. These variants are respectively called PRP-F and PRP-V. The Picker Routing Problem in Scattered Storage warehouses (PRP-SS) is another variant worth mentioning, where each item may be available in several locations. The PRP-SS consists in jointly routing the picker and deciding which locations to visit, it has been proven strongly NP-hard by \cite{weidinger_picker_2018}, even in single-block warehouses.

\paragraph{Complexity results for related TSP variants.}

The Traveling Salesman Problem (TSP) is a well-known NP-hard problem that consists in finding a tour of minimum weight that visits all vertices of a complete weighted graph \citep{karp_reducibility_1972}. As defined in the previous section, it is clear that the PRP is a TSP defined on a highly structured graph. The Steiner TSP (S-TSP) is an extension of the TSP, where only a subset of vertices must be visited. Given an undirected weighted graph $\mathcal{G} = (\mathcal{V},\mathcal{E})$, which may not necessarily be complete, along with a subset $\mathcal{V}' \subset \mathcal{V}$ of vertices, the S-TSP consists in finding a minimum-weight tour that visits each vertex of $\mathcal{V}'$ exactly once. The optional vertices $\mathcal{V} \backslash \mathcal{V}'$ are called Steiner points. Indeed, \cite{scholz_new_2016} show that the PRP can be seen as an S-TSP, with the red diamonds in Figure~\ref{fig:layout_2blocks} representing the Steiner points. The work of \cite{cambazard_fixed-parameter_2018} relies on the S-TSP and Steiner tree literature to propose a fixed-parameter algorithm for the Rectilinear TSP (R-TSP), which serves as a basis for recent advances on the PRP \citep{pansart_exact_2018}. The R-TSP is a variant of the TSP where the points to visit are located in the plane, and the Manhattan distance is used to calculate the distance between two points. The R-TSP is known to be NP-hard \citep{cambazard_fixed-parameter_2018,itai_hamilton_1982}. 
Note that the R-TSP is fairly close to the PRP. However, there are two main differences between the R-TSP and the PRP: {\em i.} In the PRP, the distances between two vertices are not computed using a conventional p-norm, and {\em ii.} The PRP is considerably more structured than the R-TSP, as the points in the PRP are located on a discrete grid rather than in $\mathbb{R}^2$.

\paragraph{Complexity results for related Hamiltonian cycle problems.}

The Hamiltonian Cycle Problem (HCP) is one of the well-known Karp's 21 NP-complete problems \citep{karp_reducibility_1972}. The HCP aims to decide whether an undirected non-weighted graph admits a cycle that visits each vertex exactly once. The HCP variant of interest within this note is the Hamiltonian Cycle Problem on Grid Graphs (HCPGG), defined as follows. Consider $\mathcal{G}^{\infty}$ the infinite graph whose vertices are the points of the plane with coordinates in $\mathbb{Z}^2$. Vertices in this graph  are connected if and only if the Euclidean distance between them is equal to 1. A {\em grid graph} is a subgraph of $\mathcal{G}^\infty$ induced by a finite number of vertices. Figure~\ref{fig:grid_graph:1} illustrates an example of a grid graph. The HCPGG asks whether a Hamiltonian cycle exists within a given grid graph. It is worth noting that \cite{itai_hamilton_1982} have proven the NP-completeness of the HCPGG. In the next section, we use this result to establish the intractability of the PRP. It is important to emphasize that an HCPGG instance is fully characterized by the set of vertices $\mathcal{V}^{HCPGG}$, represented by their coordinates in the plane.


\section{Complexity of the PRP in conventional warehouses}\label{sec:complexity}

Let us introduce the main result of this technical note.

\begin{theorem}\label{theorem:complexity}
    The Picker Routing Problem in multi-block warehouses is strongly NP-hard.
\end{theorem}

Before proving Theorem~\ref{theorem:complexity}, we first introduce the following lemmas that establish the complexity of the HCPGG and its restriction to connected graphs.

\begin{lemma}\label{lemma:1}
    The HCPGG is NP-complete in the strong sense.
\end{lemma}

\paragraph{Proof of Lemma~\ref{lemma:1}.}
\cite{itai_hamilton_1982} have proven that the HCPGG is NP-complete, prompting the need to further demonstrate its NP-completeness in a strong sense. In \cite{garey_computers_1979}, p.94--95, the authors define a {\em number problem} as a decision problem where the numerical values of an instance cannot be polynomially bounded by the length of the instance itself. Furthermore, according to the authors, {\em \lq\lq Assuming that P $\neq$ NP, the only NP-complete problems that are potential candidates for being solved by pseudo-polynomial time algorithms are those that are number problems. [...] In particular, if $\Pi$ is NP-complete and $\Pi$ is not a number problem, then $\Pi$ is automatically NP-complete in the strong sense.\rq\rq} Our goal is thus to show that the HCPGG is not a number problem, which is sufficient to conclude the proof.

An instance of the HCPGG is represented by a grid graph $\mathcal{G}^{HCPGG} = (\mathcal{V}^{HCPGG},\mathcal{E}^{HCPGG})$, which is fully defined by its set of vertices \citep{itai_hamilton_1982}. If we encode the vertices using their coordinates, then the numerical values of these coordinates are not polynomially bounded by $|\mathcal{V}^{HCPGG}|$. However, if we encode $\mathcal{G}^{HCPGG}$ by its adjacency matrix, which is a boolean matrix of size $|\mathcal{V}^{HCPGG}|^2$ that contains $1$ in position $(i,j)$ if vertex $v^i$ is adjacent to vertex $v^j$, and $0$ otherwise, the numerical values of the instance are bounded by a polynomial of the length of the problem. Therefore, the HCPGG is not a number problem, from which it follows that the HCPGG is NP-complete in the strong sense. This concludes the proof.
\begin{flushright}
    $\square$
\end{flushright}

\begin{lemma}\label{lemma:2}
    The HCPGG restricted to connected graphs is NP-complete in the strong sense.
\end{lemma}

\paragraph{Proof of Lemma~\ref{lemma:2}.}
We will prove this result with a straightforward reduction from the HCPGG as follows. Let $\mathcal{I}^{HCPGG}$ be an instance of the HCPGG represented by the grid graph $\mathcal{G}^{HCPGG}$. First, the reduction determines if $\mathcal{G}^{HCPGG}$ is connected. This step can be performed in polynomial time with Prim's algorithm \citep{prim_shortest_1957}. Depending on this answer, two cases appear:

\begin{itemize}
    \item \textit{$\mathcal{G}^{HCPGG}$ is connected.} Then $\mathcal{I}^{HCPGG}$ is a valid instance for the problem restricted to a connected graph, whose answer remains unchanged.
    \item \textit{$\mathcal{G}^{HCPGG}$ is not connected.} Then there cannot exist a cycle in $\mathcal{G}^{HCPGG}$, and $\mathcal{I}^{HCPGG}$ has a \lq\lq NO\rq\rq\ answer. It is sufficient for the reduction to build any connected grid graph of polynomially equivalent size that does not admit a Hamiltonian cycle.
\end{itemize}

This reduction is clearly polynomial, which concludes the proof.
\begin{flushright}
    $\square$
\end{flushright}

\paragraph{Proof of Theorem~\ref{theorem:complexity}.}
To prove that the Picker Routing Problem (PRP) is NP-hard in the strong sense, we will exhibit a polynomial reduction from the HCPGG to the decision version of the PRP, i.e., the problem that decides whether a PRP instance admits a solution of value lower than or equal to a given quantity $t \in \mathbb{R}$.
Let $\mathcal{I}^{HCPGG}$ be an instance of the HCPGG represented by graph $\mathcal{G}^{HCPGG} = (\mathcal{V}^{HCPGG},\mathcal{E}^{HCPGG})$. According to Lemma~\ref{lemma:2}, we can assume, without loss of generality, that $\mathcal{G}^{HCPGG}$ is connected. For a vertex $v \in \mathcal{V}^{HCPGG}$, we denote its coordinates in the plane as $(v_x,v_y) \in \mathbb{Z}^2$. We will reduce $\mathcal{I}^{HCPGG}$ to the decision version of an instance of the PRP. Since $\mathcal{G}^{HCPGG}$ is defined over a finite number of vertices, there exists $n,m \in \mathbb{N}^2$ such that $\mathcal{G}$ is included in a rectangular grid graph $\mathcal{G}^*$ of size $n \times m$. In other words, $\mathcal{G}^*$ is induced by the vertices $[\alpha, \, \alpha + n] \times [\beta , \beta + m]$ for some $\alpha, \beta \in \mathbb{Z}^2$. Without loss of generality, we can assume that $\alpha = \beta = 0$. Furthermore, we can assume that $\mathcal{G}^*$ is tight on the bottom of the grid, meaning there exists a vertex $v^{bottom} \in \mathcal{V}^{HCPGG}$ with its y-coordinate equal to 1 ($v^{bottom}_y = 1$). Similarly, we assume that $\mathcal{G}^*$ is tight on the left of the grid. Since $\mathcal{G}^{HCPGG}$ is connected and $\mathcal{G}^*$ is tight on the bottom and left, it appears that $n$ and $m$ can both be bounded by the number of vertices $|\mathcal{V}^{HCPGG}|$.

\begin{figure}[h]
    \centering
    \begin{minipage}[b]{0.45\textwidth}
        \includegraphics[width=\textwidth]{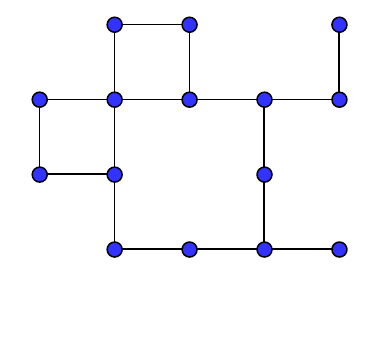}
        \caption{Example of a grid graph.}
        \label{fig:grid_graph:1}
    \end{minipage}
    \hfill
    \begin{minipage}[b]{0.45\textwidth}
        \includegraphics[width=.9\textwidth]{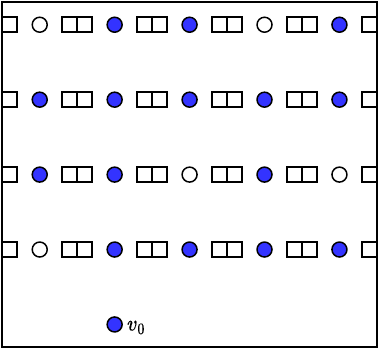}
        \caption{PRP instance after reduction (only blue locations need to be visited).}
        \label{fig:grid_graph:2}
    \end{minipage}    
\end{figure}

Let us consider a rectangular warehouse composed of $m$ blocks, $n$ aisles, and a single location per aisle. In this case, for $k \in \mathcal{K}$ and $a \in \mathcal{A}$, the corresponding set of locations $\mathcal{L}^{ka}$ is a singleton. We denote this set $\mathcal{L}^{ka} = \{l^{ka}\}$. Let us suppose that the depot is located in front of the aisle $a^{depot} = v^{bottom}_y$ and is separated from the first cross aisle by a distance $D^{depot} = 0$. Moreover, consider that the distances between consecutive locations, aisles and blocks are set to $D^{loc} = 0$, $D^{aisle} = 1$, and $D^{block} = 1$ respectively. We define the set $\mathcal{V}^{PRP}$ of visited locations as follows. For each vertex $v = (v_x,v_y) \in \mathcal{V}^{HCPGG}$, we insert in $\mathcal{V}^{PRP}$ the only location $l^{v_yv_x}$ of block $v_y$, aisle $v_x$. The depot $v_0$ is also inserted in $\mathcal{V}^{PRP}$. More formally, we have $\mathcal{V}^{PRP} =  \{v_0\} \cup \{l^{v_yv_x}, \, \forall (v_x,v_y) \in \mathcal{V}^{HCPGG}\}$. An instance $\mathcal{I}^{PRP}$ of the PRP is defined on the complete weighted graph $\mathcal{G}^{PRP} = (\mathcal{V}^{PRP},\mathcal{E}^{PRP})$, where the weight of an edge $(v^1,v^2) \in \mathcal{E}^{PRP}$ equals the shortest walking distance between $v^1$ and $v^2$ based on the defined layout. Figure~\ref{fig:grid_graph:2} illustrates this transformation. It is important to note that $\mathcal{G}^{PRP}$ is complete, unlike $\mathcal{G}^{HCPGG}$. Additionally, we observe that an edge of $\mathcal{E}^{HCPGG}$ corresponds to an edge of weight 1 in $\mathcal{E}^{PRP}$, while any other edge of $\mathcal{E}^{PRP}$ either {\em i.} visits the depot, or {\em ii.} has a weight greater than or equal to 2.

Let us now prove that $\mathcal{I}^{HCPGG}$ has a \lq\lq YES\rq\rq\ answer (i.e., there exists a Hamiltonian cycle in $\mathcal{G}^{HCPGG}$) if, and only if, the corresponding $\mathcal{I}^{PRP}$ admits a solution with a value of at most $|\mathcal{V}^{HCPGG}|$. 

Let us first suppose there is a Hamiltonian cycle in $\mathcal{G}^{HCPGG}$, then its equivalent in $\mathcal{G}^{PRP}$ forms a cycle of value $|\mathcal{V}^{HCPGG}|$. We need, however, to add a visit to the depot to get a valid PRP solution. Since we built the instance such that the depot $v_0$ is located at a distance $0$ from a location $v$ that must be visited, we can insert $v_0$ just before $v$ in the cycle without altering its cost. Therefore, we built a valid PRP solution of cost $|\mathcal{V}^{HCPGG}|$.

For the converse implication, let us suppose there exists a PRP solution with a value of $|\mathcal{V}^{HCPGG}| = |\mathcal{V}^{PRP}| - 1$, i.e., a cycle that visits all vertices in $\mathcal{V}^{PRP}$, including the depot. We recall there is a single edge of cost $0$ in $\mathcal{G}^{PRP}$, that is the one connecting the depot to the closest picking location, and all the other edges have a cost greater than or equal to $1$. Given that the PRP cycle contains $|\mathcal{V}^{HCPGG}|+1$ edges, it necessarily contains the edge with cost $0$. Consequently, by removing the depot vertex from the solution, we obtain a cycle of value $|\mathcal{V}^{HCPGG}|$ that does not visit $v^0$, and thus contains only edges of value 1, i.e., edges from $\mathcal{E}^{HCPGG}$ by transformation, which exhibit a Hamiltonian cycle in $\mathcal{G}^{HCPGG}$.

We have built a reduction from the HCPGG to the PRP. Clearly, this reduction can be computed in polynomial time. Furthermore, we observe that $D^{depot} = D^{loc} = 0$, $D^{block} = D^{aisle} = L = 1$ and $A,\, K \leq |\mathcal{V}^{HCPGG}|$, so all numerical values of $\mathcal{I}^{PRP}$ are bounded by the length of the problem. Therefore, the reduction is polynomial and the PRP is NP-hard in the strong sense.
\begin{flushright}
    $\square$
\end{flushright}

It is important to note that, without the connectivity assumption provided by Lemma~\ref{lemma:2}, the reduction exhibited in the proof of Theorem~\ref{theorem:complexity} would not be polynomial anymore, since it would exponentially increase the numerical values in the constructed PRP instance.


\section{Complexity results for related problems}\label{sec:additional}

%

While the complexity of most problems in the OP literature is already well established, there remains some variants whose complexity status is still open. In this section, we use Theorem~\ref{theorem:complexity} to close some of these gaps. The validity of these results is straightforward as the PRP is a special case of the problems, hence we omit the proofs in this section.

\paragraph{The Order Batching Problem (OBP)}
The integration of the PRP with the \textit{Order Batching Problem}, which consists in batching orders together with limited capacity such that the total distance is minimized, was studied by \cite{gademann_order_2005} in a single-block warehouse. They proved that the problem is strongly NP-hard when the picking tours may contain more than two orders, while it is polynomially solvable otherwise. While NP-hardness result remains valid for the multi-block case in general, the polynomial algorithm the authors exhibit when each batch contains at most two orders is only valid for the single-block layout. The proof of \cite{gademann_order_2005} relies on a polynomial oracle for computing picking route lengths, which renders it invalid for the multi-block case.

\begin{corollary}\label{cor1}
    The Order Batching Problem in multi-block warehouses is NP-hard in the strong sense, even when each batch contains at most two orders.
\end{corollary}

\paragraph{The prize collecting PRP}
\cite{bock_routing_2023} study the prize collecting variant of the PRP in multi-block warehouses. They show that the problem is NP-hard in the general case, and admits a pseudo-polynomial algorithm when the number of blocks $K$ satisfies $K \in \mathcal{O}(\log n)$. This establishes the problem as weakly NP-hard under this condition. However, the complexity status for the general case remains a open question. By using theorem~\ref{theorem:complexity}, we can definitively classify the problem.

\begin{corollary}\label{cor2}
    The prize-collecting PRP is strongly NP-hard in multi-block warehouses.
\end{corollary}

\paragraph{Other integrated order picking problems}
More generally, the result of Theorem~\ref{theorem:complexity} implies the strong NP-hardness of all problems where one of the decisions is the routing of pickers. In addition to the new findings presented in Corollaries~\ref{cor1} and~\ref{cor2}, the present note provides an alternative and straightforward complexity proof for a wide range of integrated order picking problems within the context of multi-block warehouses.

\begin{corollary}
    Any planning problem involving picker routing decisions is strongly NP-hard in multi-block warehouses.
\end{corollary}


\section{Conclusion}\label{sec:conclusion}

In this note, we have shown that the Picker Routing Problem (PRP) in a rectangular warehouse with parallel aisles is NP-hard in the strong sense. This result provides a rationale for the development of heuristic methods for solving the PRP, despite the efficiency of exact algorithms. Indeed, in practice, the fixed-parameter tractability of the PRP makes the problem efficiently solvable by exact algorithms for all realistic instances \citep{pansart_exact_2018,schiffer_optimal_2022}. However, there remain time-critical applications where the development of very fast heuristic methods is of interest. For instance, this is relevant when the PRP is used as a subroutine within decomposition methods for integrated warehouse planning problems \citep{van_gils_designing_2018,briant_efficient_2020,prunet_storage_2024}.

\subsection*{Acknowledgments}

This work has been supported by the French National Research Agency through the AGIRE project under the grant ANR-19-CE10-0014\footnote{\url{https://anr.fr/Projet-ANR-19-CE10-0014}}. This support is gratefully acknowledged.




\bibliographystyle{apalike} 
\bibliography{references}

\end{document}